# An exact scalar field inflationary cosmological model which solves Cosmological constant problem and dark matter problem in addition to Horizon and Flatness problems and other problems of inflationary cosmology.

Debasis Biswas
Netaji Park, Chakdaha, Nadia, 741222, India
E Mail: biswasdebasis38@gmail.com

[Abstract: An exact scalar field cosmological model is constructed from the exact solution of the field equations. The solutions are exact and no approximation like slow roll is used. The model gives inflation, solves horizon and flatness problems. The model also gives a satisfactory estimate of present vacuum energy density and solves cosmological constant problem of 120 orders of magnitude discrepancy of vacuum energy density. Further, this model predicts existence of dark matter/energy and gives an extremely accurate estimate of present energy density of dark matter and energy. Alongwith explanations of graceful exit, radiation era, matter domination, this model also indicates the reason for present accelerated state of the universe. In this work a method is shown following which one can construct an infinite number of exact scalar field inflationary cosmological models.]

## 1. Introduction.

Inflation was proposed by Alan Guth [1] although the idea of an exponential type expansion was due to Starobinsky and others [2,3,4]. The modern form of inflationary cosmology is due to A.Linde, A. Albrecht and P. Steinhardt [5, 6]. In Guth's original model the inflaton field $\Phi$ was assumed to be trapped in a false vacuum and assumed a local value which is minimum. The inflaton field comes out from the local minimum value by quantum tunnelling and as universe inflates, tunnelling takes place. However, these ideas when pursued gave empty universe and therefore rejected. Guth further tried to improve the idea but they led to others difficulties.

Linde and Steinhardt proposed new inflationary model where the inflaton field varies slowly and undergoes a phase transition of second order. New inflationary models do not require the idea of tunnelling. Most of the modern models depend on the idea of chaotic inflation due to Linde. In these models the initial value of the inflaton field $\Phi$ is set chaotically when the universe exits from Planck era. The field then rolls downhill and if the potential is enough flat then inflation can take place.

There are another class of models known as hybrid inflationary models in which two fields are considered. These models introduce extra difficulties but they can speculate some features of single field models.

Inflationary cosmology is important because it offers solution to some great puzzles of cosmology. The puzzles are Flatness problem, Horizon problem and Monopole problem.

Flatness problem is basically why the density parameter $\Omega$ (t) = $\rho/\rho_c$ is extremely close to unity i.e. why $\Omega \approx 1$? Horizon problem is why the universe is extremely smooth and isotropic on large scales ? Monopole and the unwanted relics are the problems associated with standard hot Big Bang Theory. They are trivially solved when Flatness and Horizon problems are solved.

The above problems namely Flatness problem and Horizon problem are problems of Standard Big Bang theory are solved by assuming an accelerated expansion in early universe for a very short duration. This accelerated expansion is named as inflation. The starting time of inflation is model dependent. However, it occurred when the universe was extremely young. Inflation ended around the time when universe was $10^{-33}$sec.old. From this time ($10^{-33}sec.$)radiation domination started. The phenomenon of ending inflation and then entering into radiation dominated era is known as graceful exit. And its mechanism requires explanations. An entirely different mechanism of graceful exit will be given in this work.

Lot of scalar field inflationary cosmological models have been proposed so far to explain the above scenarios. Expansion of universe is assumed to be driven by a scalar field $\mathbf{\Phi}$ and an associated potential $\mathbf{V(\Phi)}$**.** Many forms of potentials [7, 8, 9, 10, and 11] have been used to solve the associated field equations.

In some models a kind of approximation is used to solve the difficult equations. This approximation is known as slow roll approximation which assumes that the field rolls very slowly. Mathematically this is equivalent to assuming $\dot{\mathbf{\Phi}}^{\mathbf{2}} \ll \mathbf{V(\Phi)}$ where the overhead dot represents derivative with respect to time. A few models find exact solutions to the field equations. All the above models explain the mechanism of inflation and solve Horizon and Flatness problems. Further it is found that solution of these problems are equivalent to produce an e-folding [defined as $\mathbf{ln\frac{a_f}{a_i}}$ during inflation] **N ≥ 65-70** [12]. Here $\mathbf{a_i}$ and $\mathbf{a_f}$ are values of scale factor when inflation starts and ends respectively.

However these above models fail to explain cosmological constant problem [13] and dark matter problem. The cosmological constant problem is why the measured vacuum energy density is small by a factor of about $\mathbf{10^{120}}$ from its theoretical value. This is in language of Weinberg; "Worst failure of an order of magnitude estimate in the history of physics".

The dark matter problem is another unsolved puzzle in modern cosmology. Our present knowledge asserts that the energy density of matter/energy content [14] of our universe is: dark energy ~ 74%, dark matter ~ 22% and ordinary matter ~ 4%. No cosmological model predicts or accounts for this observation.

Finally, there is the problem of present acceleration [15, 16, 17] of the universe found from the observation of distant Supernovae Ia.

The present work addresses all the above problems listed from the beginning and provides solutions in a single framework. Further, the solution of cosmological evolution equations are exact and no sort of approximations like slow roll approximation etc. is used to derive the solutions.

It may be mentioned here that slow roll is not the necessary and sufficient condition of inflation. However, if slow roll is valid, inflation takes place. It will be shown in this work that without slow roll one can have plenty of exact inflationary models.

## 2. The scalar field equation and its exact solutions.

We suppose that after tunnelling there exists a scalar field $\mathbf{\Phi}$ and an associated potential $\mathbf{V(\Phi)}$**,** which is responsible for the evolution of the universe. It is further assumed that initially there existed some other type of fields $\mathbf{\Psi_i}$ with potentials $\mathbf{X(\Psi_i)}$**.** But these fields were hanged up initially which means $\dot{\Psi}_i(\frac{d\Psi_i}{dt})$ and $X'(\Psi_i)$ [$=\frac{dX}{d\Psi_i}$] are negligible and they did not contribute to field equations initially. The number and nature of the $\Psi_i$fields are not important for the purpose of cosmological predictions. The interactions of the scalar field $\Phi$ with other fields are assumed to be ignorable and consequently the $\Psi_i$ fields are assumed to interact among themselves only.

Now if the inflaton field $\Phi$ has no spatial variation and depends only on time then we can write the equations of motion [18] of the scalar field and the Friedmann equation ignoring the curvature term as:

$$\ddot{\Phi} + 3\frac{\dot{a}}{a}\dot{\Phi} + V'(\Phi) = 0 \qquad (1)$$

and

$$\frac{\dot{a}^2}{a^2} = \frac{1}{3}\left[\frac{1}{2}\dot{\Phi}^2 + V(\Phi)\right] \qquad (2)$$

$$= \frac{1}{3}\rho_\Phi \qquad (2a)$$

Where **a** is the scale factor, $\Phi$ is the inflaton field and V ($\Phi$) is the potential. Overhead dot represents derivative with respect to time and overhead prime represents derivative w.r.to $\Phi$.

Equation (1) follows from the Lagrangian [18]

$$L = \frac{1}{2}g^{\mu\nu}(\partial_\mu\Phi)(\partial_\nu\Phi) - V(\Phi) \qquad (3)$$

Solution of equation (1) and (2) are in some ways similar to the solution of Diophantine equations in Classical Algebra [19], where the number of unknowns are more than the number of equations given.

Here a method will be shown by which one can find exact solution of equation (1) and (2). In principle we will choose an arbitrary function from which we can construct some form of potentials for which equations (1) and (2) are exactly solvable.

Following this method [Appendix A] one can find as many as exact solutions as one wishes. (In principle this method allows one to find an infinite number of exact solutions.)

Now following the method derived and illustrated in Appendix-A, we write the solutions of (1) and (2).

They are:

$$\frac{\dot{a}}{a} = \frac{1}{\sqrt{6}}f(t)^{1/2} \qquad (4)$$

$$V(\Phi) = w(t) = \frac{f(t)}{2} + \frac{\dot{f}(t)}{2\sqrt{6}f(t)^{1/2}} \qquad (5)$$

And $\dot{\Phi}^2 = -\frac{1}{\sqrt{6}}\frac{\dot{f}(t)}{f(t)^{1/2}}$ (6) (The overhead dot represents time derivative.)

The functions $f(t)$ is arbitrary so that one can have an infinite number of choices of $f(t)$ and can have an infinite number of exact solutions.

## 3. The exact scalar field model and solution of Flatness and Horizon problems.

From the method illustrated in Appendix-A we can now find an exact inflationary model.

We choose the arbitrary function:

$$f(t) = \left(\frac{A}{t} + B\right)^2 \qquad (7)$$

The results are [Appendix A]

$$a = A_0 e^{\frac{+B}{\sqrt{6}}t} t^{A/\sqrt{6}} \qquad (8)$$

Where $A_0$, A and are real arbitrary constants.

For A> $\sqrt{6}$ (of course $A_0 > 0$), and B>0, one can observe that ä> 0 always. Therefore the above scale factor gives inflation. We will choose later on A such that A> $\sqrt{6}$ and B>0.

The potential [Appendix-A] which gives the above scale factor is in time-dependent form [A.29]:

$$w(t) = \left(\frac{A^2}{2} - \frac{A}{\sqrt{6}}\right)\frac{1}{t^2} + \frac{AB}{t} + \frac{B^2}{2} \qquad (9)$$

And the same potential [A.34] in $\Phi$ dependent form is:

$$V(\Phi) = Ce^{2\Phi/A_1} + De^{\Phi/A_1} + \frac{B^2}{2} \qquad (10)$$

In this model we choose the starting time of inflation as $t_i = 10^{-43}$ second i.e. just after tunneling and inflation ends at $t_f = 10^{-33}$second. Particle production in inflaltionary period is assumed to be negligible and ignored.

The mechanism of ending inflation will be discussed in next section.

Now from equation (8) we can find the e-folding during inflation.

$$N = \ln\frac{a_f}{a_i} = \ln\left\{e^{\frac{B}{\sqrt{6}}(t_f - t_i)}\right\} + \frac{A}{\sqrt{6}}\ln\left(\frac{t_f}{t_i}\right) \qquad \text{using (8)}$$

$$i.e. \quad N = \frac{B}{\sqrt{6}}(t_f - t_i) + \frac{A}{\sqrt{6}}\ln\left(\frac{t_f}{t_i}\right) \qquad (11)$$

Now we take $B = 10^{-17}\,\text{sec.}^{-1}$ (11a) [for inflation to take place A>$\sqrt{6}$ and B>0]

And $\left.\begin{array}{l}\mathbf{t_f = 10^{-33}sec.}\\ \mathbf{t_i = 10^{-43}sec.}\end{array}\right\}$ (11b)

Then using (11a) and (11b) we obtain from (11)

$$\mathbf{N \sim \frac{10A}{\sqrt{6}} \ln 10}$$

$$\mathbf{= \frac{10A \times 2.3025}{2.4494}}$$

i.e. $\mathbf{N = 9.40A}$ (12)

For inflation to take place **A>**$\sqrt{\mathbf{6}}$

If we choose A=7.5

Then from (12) the result is, $\mathbf{N = 9.40 \times 7.5}$

i.e. $\mathbf{N = 70.5}$ (13)

Therefore the e-folding one obtains is 70.5, which is perfectly satisfactory.

**4.** Graceful exit and starting of radiation era.

It was assumed in previous section that inflation starts at $\mathbf{t_i = 10^{-43}}$sec. and stops at $\mathbf{t_f = 10^{-33}sec}$. The mechanism by which inflation stops is like this. It was postulated in section 2, that there were some hanged up fields for which $\dot{\mathbf{\Psi}}_{\mathbf{i}}$ and $\mathbf{X}'(\mathbf{\Psi_i})$ were negligible so that they did not contribute to the field equations. When inflation starts the inflaton field decays. During the period of inflation particle production due to decaying inflaton field is assumed to be negligible and not taken into account. But all of the hanged up fields interact among themselves and produce new particles with significant negative energy density around the time $\mathbf{10^{-34}sec}$. The effect of these negative energy density particles is to stop inflation at $\mathbf{t_f = 10^{-33}sec}$. The newly born fields created by these particles are denoted by $\mathbf{\Psi_{E_i}}$.

The equation of state of dark energy [12] is $\boldsymbol{\omega} = \mathbf{P}/_{\rho}$ where **ω ≈ -1**. For dark matter we assume the same equation of state as dark energy but a different negative value of **ω**. Now since **ω** is negative,there exist two possibilities **i) P>0, ρ<o** or ii) **P<0, *ρ*>0**. Generally for dark energy the second possibility is accepted.

High energy physics assert that many forms of exotic particles form around the time $\mathbf{t = 10^{-34}}$second. The natures of the particles depend on the theory concerned and their natures are not very important for our purpose. We take it for granted that many forms of exotic particles are formed around the time $\mathbf{10^{-34}sec}$. In analogy with dark energy equations of state we take the equation of state of these particles as $\boldsymbol{\omega} = \mathbf{P}/_{\rho}$ with **ω** negative. However, we take the first possibility discussed before i.e. we take **P>0** and **ρ<0** for these exotic particles. And appearance of a large negative energy density field is enough to stop inflation at $\mathbf{t_f = 10^{-33}}$sec. Because creation of a large number of exotic particles with properties **P>0** and **ρ<o** will certainly decrease the energy density and create a situation for which an overall condition **3P + ρ > 0** would appear if we take $\mathbf{-1 \leq}$ **ω < -** $\mathbf{\frac{1}{3}}$ for these particles, as it turns out that **3P + ρ > 0** for these large number of exotic particles**.** As a result inflation must stop.

The assumption of negative energy density particles is perfectly consistent with the Null energy condition and Strong energy condition [13].

The appearance of new negative energy density due to creation of new particles does not alter equation (1) though they contribute to the Lagrangian from this time $(\mathbf{t \sim 10^{-33} sec.})$**.** The reasons are, the inflaton field $\mathbf{\Phi}$ has no appreciable interactions with the $\mathbf{\Psi}_i$ fields or with the newly born $\mathbf{\Psi_{E_i}}$ fields at the time of graceful exit.

But the Friedmann equation assumes a new form from the time of graceful exit. Considering the appearance of negative energy density particles we find that Friedmann equation **[** i.e. equation (2a) **]** assumes its new form at the time of graceful exit :

$$\frac{\dot{a}^2}{a^2} = \frac{1}{3}\left(\rho_\Phi - \sum \rho_{Ei}\right) \qquad (14)$$

Where $\sum \boldsymbol{\rho}_{\mathbf{Ei}} = \sum \frac{1}{2} \dot{\mathbf{\Psi}}^2{}_{\mathbf{Ei}} + \sum \mathbf{V}\ (\mathbf{\Psi}_{\mathbf{E_i}})$ **.**

We neglect further variations of $\dot{\mathbf{\Psi}}_{\mathbf{E_i}}$ and $\mathbf{V}(\mathbf{\Psi}_{\mathbf{E_i}})$ and they do not interact further among themselves.

Here $\sum \boldsymbol{\rho}_{\mathbf{Ei}}$ is the energy density of the new exotic particles formed. The negative sign indicates that the energy densities of the exotic particles are negative.

At the time of graceful exit the universe enters into a decelerated phase. It is known [18] that the conditions of accelerated phase is **3P+*ρ*<0** and that of decelerated phase is **3P+ρ>0**. We can therefore assume that the creation of new energy density due to newly born particles create an overall situation where an overall condition like **3P+ρ>0** holds from the time of graceful exit.

The foregoing discussions illustrate the mechanism of graceful exit. An accelerated expansion reduces to a time half power law at the time of graceful exit i.e. at $\mathbf{t = 10^{-33} second}$. So from this time radiation era starts.We can exactly calculate value of $\sum \boldsymbol{\rho}_{\mathbf{Ei}}$ at the time of graceful exit using (A.29a) and (A.30) and taking $\mathbf{a \sim t^{1/2}}$**.**This is, however unnecessary for our purpose.

## 5. Cosmological constant and dark matter/energy problem.

After graceful exit the expansion of universe continues and the inflaton field $\mathbf{\Phi}$ goes on decaying. The energy density gradually increases. We assume that particles are produced in this phase with properties **P**$\geq$ **0** , **ρ>o** as well as **P<0 ρ>o** . For the second type of particles if we assume an equation of state $\boldsymbol{\omega} = \mathbf{P}/_{\rho}$ $\mathbf{with}\ -\mathbf{1} \leq \boldsymbol{\omega} < -\frac{\mathbf{1}}{\mathbf{3}},$ then **3P** + **ρ** **< 0** for these particles. All energy conditions permit this [13]. We take it for granted that these type of particles are produced more than the first type in matter dominated phase. Now $\sum$ **ρ > 0** , since **ρ > 0** for both type of particles. The overall effect is the appearance of a positive energy density denoted by $\sum \boldsymbol{\rho}_{\mathbf{I}}$ .Thus total energy density of all created particles after graceful exit upto present moment is represented by $\sum \boldsymbol{\rho}_{\mathbf{I}}$ .

With this idea we can now write the Friedmann equation at present epoch:

$$\frac{\dot{a}^2}{a^2} = \frac{1}{3}\left[\rho_\Phi + \Sigma\rho_I - \Sigma\rho_{Ei}\right] \qquad (15)$$

Equation (15) follows from (14) by introducing the term $\sum \rho_I$ in R.H.S of (14)

Here $\rho_\Phi$ is the energy density of the inflaton field.

i.e. $\rho_\Phi = \frac{1}{2}\dot{\Phi}^2 + V(\Phi)$

And $\sum \rho_I$= energy density of the created particles after graceful exit upto present epoch.

And $\Sigma\rho_{Ei}$ = energy density of exotic particles created just before the time of graceful exit.

It is difficult to calculate $\sum \rho_I$ but one can safely assume that $\sum \rho_I$ is much less than $\Sigma\rho_{Ei}$ , so that we can write:

$$\sum \rho_I - \Sigma\rho_{Ei} = -\sum \rho_* \quad (16)$$

Then equation (15) can be recasted as

$$\rho_\Phi - \sum \rho_* = \frac{3\dot{a}^2}{a^2} = 3H_\circ^2 \quad (17)$$

Where $H_\circ$ is the present Value of Hubble constant.

Using the present value of $H_\circ$[18] as

$$H_\circ = 2.27\times10^{-18}\text{sec}^{-1} \quad (18)$$

We find from (17) the present value of $\rho_\Phi - \sum \rho_*$ as

$$\rho_\Phi - \sum \rho_* = 3\times(2.27)^2 \times 10^{-36} = 1.54 \times 10^{-35}\text{sec.}^{-2} \quad (19)$$

Now we define cosmological constant as the energy density of the inflaton field (i.e. $\rho_\Phi$) as

$$\rho_\Phi = \frac{1}{2}\dot{\Phi}^2 + V(\Phi) \quad (20)$$

Using (A.29 a) and (A.29) we write

$$\rho_\Phi = \frac{1}{\sqrt{6}}\frac{A}{t^2} + \left(\frac{A^2}{2} - \frac{A}{\sqrt{6}}\right)\frac{1}{t^2} + \frac{AB}{t} + \frac{B^2}{2}$$

i.e. $\boldsymbol{\rho_\Phi} = \frac{A^2}{2t^2} + \frac{AB}{t} + \frac{B^2}{2}$ (21)

Now taking A=7.5 and B=$10^{-17}$ as earlier, we find

$$\boldsymbol{\rho_\Phi} = \frac{28.12}{t^2} + \frac{7.5x10^{-17}}{t} + 0.5\times10^{-34} \quad (22)$$

Then from (22) at $\mathbf{t = 10^{-43}sec}$. i.e. at Planck epoch,

$$\boldsymbol{\rho_\Phi(t = 10^{-43}sec.) = 28.12 \times 10^{86} + 7.5 \times 10^{26} + 0.5 \times 10^{-34}}$$
$$\boldsymbol{\approx 2.81 \times 10^{87} sec^{-2}} \quad (23)$$

And at present i.e. at $\mathbf{t = 4.4 \times 10^{17} sec}$.

$$\boldsymbol{\rho_\Phi(t = 4.4 \times 10^{17} sec.) = \frac{28.12}{(4.4)^2} \times 10^{-34} + \frac{7.5 \times 10^{-34}}{4.4} + 0.5 \times 10^{-34}}$$

$$\boldsymbol{= 1.45 \times 10^{-34} + 1.70 \times 10^{-34} + 0.5 \times 10^{-34}}$$

$$\boldsymbol{= 3.65 \times 10^{-34} sec.^{-2}} \quad (24)$$

Then using (23) and (24)

$$\frac{\boldsymbol{\rho_\Phi}(t=10^{-43}Sec)}{\boldsymbol{\rho_\Phi}(t=4.4\times10^{17}Sec)} = \frac{2.81\times10^{87}}{3.65\times10^{-34}} = 7.69 \times 10^{120} \quad (25)$$

Equation (24) gives the present value of cosmological constant and equation (25) exactly accounts for the so called discrepancy of 120 orders of magnitude of the value of cosmological constant.

Since L.H.S of (17) represents effective vacuum energy density at present, so more precise present value of cosmological constant is given by (19) and equals $\mathbf{1.54 \times 10^{-35} sec.^{-2}}$. Then using this value we find from (23)

$$\frac{\boldsymbol{\rho_\Phi}(t=10^{-43}Sec)}{\boldsymbol{\rho_\Phi}(\text{present})} = \frac{2.81\times10^{87}}{1.54\times10^{-35}} = 1.82 \times 10^{122} \quad (25a)$$

Equation (25a) gives more precise ratio of cosmological constant at Planck epoch and at present epoch.

Now we indentify $\sum \boldsymbol{\rho}_*$defined by equation (16) is the energy density of dark matter/energy and calculate its present value. The negative sign before $\sum \boldsymbol{\rho}_*$ in (16) indicates that energy density of dark matter/ energy is negative.

Using (19) and (24) we find the present value of energy density of dark matter/energy as

$$\sum\boldsymbol{\rho}_* = \mathbf{3.65 \times 10^{-34} - 1.54 \times 10^{-35}}$$

$$= \mathbf{36.5 \times 10^{-35} - 1.54 \times 10^{-35}}$$

i.e. $\sum \boldsymbol{\rho}_* = \mathbf{34.96 \times 10^{-35} sec.^{-2}}$ (26)

Now using (24) and (26) the present ratio of $\sum \boldsymbol{\rho}_*$ and $\boldsymbol{\rho_\Phi}$ is obtained as:

$$\frac{\sum \boldsymbol{\rho}_*}{\boldsymbol{\rho_\Phi}} = \frac{\mathbf{34.96 \times 10^{-35}}}{\mathbf{3.65 \times 10^{-34}}} = \frac{\mathbf{34.96}}{\mathbf{36.5}} = \mathbf{.9578} \quad (27)$$

In view of equation (27) we can safely conclude that 95.78% energy density of the inflaton field is diminished by the presence of negative energy density of dark matter/energy and the rest 4.22% represent ordinary matter energy, since for ordinary matter / energy **ρ>0** [12]. Thus the present energy density budget of the universe finds its correct accounting, 95.78% corresponds to dark matter and energy and 4.22% corresponds to ordinary matter and energy. However there is a basic difference in the nature of the above energy densities. The energy density of inflaton i.e. vacuum energy density is positive, while the energy density of dark matter/energy is negative . The present, energy density of ordinary matter-energy equals present vacuum energy density less the magnitude of present energy density of dark matter/energy. And as energy density of exotic particles were taken negative , it turns out that constituents of dark matter/energy are exotic particles as energy density of dark matter/energy is also negative.

## **6.** Matter domination and present accelerated state of the universe.

It was explained in previous sections that the mechanism of graceful exit is due to formation of some kinds of particles due to interaction of the hanged up fields themselves.

Now during the course of evolution, after graceful exit the energy density slowly increases due to further formation of new particles. Unlike exotic particles energy density, these particles have positive energy densities. So that they add up with inflaton energy density $\boldsymbol{\rho_\Phi}$. Cooling also increases of the energy density of the universe. And due to this overall increase of energy density, the universe gradually enters into matter dominated phase, when formation of matter takes place.

Present accelerated phase is due to further continuation of above features i.e. formation of more and more positive energy density particles together with cooling etc. It was assumed in section 5 that particles produced after graceful exit has the property $\sum \boldsymbol{\rho}_\mathbf{I} > 0$ and in matter dominated phase more particles are produced with property **ρ > 0, P< 0** than particles with property **ρ > 0**, **P** $\geq$ **0**. The equation of state of the particles with property **ρ > 0, P< 0** is such that **3P + ρ < 0** . Particles with **ρ > 0**, **P** $\geq$ **0** are ordinary matter /radiation, whereas particles with **$\rho > 0$, $P < 0$** along with **$3P + \rho < 0$** probably represent unstable particles which have vacuum like properties. Now in matter dominated phase as more and more particles are produced with property **$\rho > 0$, $P < 0$**, **$3P + \rho < 0$** , a situation is gradually reached for which **$\sum 3P + \sum \rho < 0$** . And acceleration of the universe starts right from the moment when **$\sum 3P + \sum \rho$** becomes negative . Such a situation still continues for which we observe our universe accelerating presently . It is once again mentioned that particles produced in various phases after graceful exit has properties **$\rho > 0$, $P < 0$** as well as **$\rho > 0$**, **$P \geq 0$ ,** whereas for exotic particles which were formed just before graceful exit **$\rho < 0$, $P > 0$ .**

## 7. Conclusion.

A variety of cosmological models were proposed in last three decades to solve the major problems of cosmology. Among these are the Coleman-Weinberg SU (5) model, models by Pi[20]and Shafi and Vilenkin [21] and many other models. All the above models were either a failure or partially successful to explain few features only. And all models so far proposed failed to explain the mysterious cosmological constant problem.

Also no model has yet predicted the existence of dark matter and energy.

The present work solves the mysterious cosmological constant problem i.e. the discrepancy of 120 or more precisely 122 orders of the measured value of cosmological constant and predicts the existence of dark matter and energy. The work removes the ambiguity of definition of cosmological constant by clearly defining it as scalar field energy density or vacuum energy density and not the energy density of dark matter/energy. Further, this model gives extremely accurate estimate of present values of vacuum energy density as well as of energy density of dark matter/energy. It also solves flatness and horizon problem, gives a satisfactory estimate of e-folding which is necessary to solve horizon and flatness problems and of course trivially monopole problem.

Finally this work also supplies the explanation for the present state of acceleration of the universe.

<u>Appendix A</u>

The Friedmann and Scalar Field equations are

$$\frac{\dot{a}^2}{a^2} = \frac{1}{3}\left(\frac{1}{2}\dot{\Phi}^2 + V(\Phi)\right) \qquad \text{(A.1)}$$

$$\ddot{\Phi} + 3\frac{\dot{a}}{a}\dot{\Phi} + V'(\Phi) = 0 \qquad \text{(A.2)}$$

Where the dots represent derivative with respect to time **t** and prime represents derivative with respect to **$\Phi$**.

From (A.1) one obtains

$$\frac{\dot{a}}{a} = \frac{1}{\sqrt{3}}\left(\frac{1}{2}\dot{\Phi}^2 + V(\Phi)\right)^{1/2} \qquad \text{(A.3)}$$

From (A.2) we have

$$\ddot{\Phi} + V'(\Phi) = -3\frac{\dot{a}}{a}\dot{\Phi}$$

$$= -3\frac{1}{\sqrt{3}}\left(\frac{1}{2}\dot{\Phi}^2 + V(\Phi)\right)^{\frac{1}{2}}\dot{\Phi} \qquad \text{using (A.3)}$$

i.e. . $\ddot{\Phi} + V'(\Phi) = -\sqrt{3}\left(\frac{1}{2}\dot{\Phi}^2 + V(\Phi)\right)^{\frac{1}{2}}\dot{\Phi}$ (A.4)

Therefore, squaring both sides of (A.4) one obtains:

$$\ddot{\Phi}^2 + V'(\Phi)^2 + 2\ddot{\Phi}V'(\Phi) = 3\left(\frac{1}{2}\dot{\Phi}^2 + V(\Phi)\right)\dot{\Phi}^2$$

After rearrangement, we have

$$\ddot{\Phi}^2 + V'(\Phi)^2 + 2\ddot{\Phi}V'(\Phi) - \frac{3}{2}\dot{\Phi}^4 - 3V(\Phi)\dot{\Phi}^2 = 0 \qquad \text{(A.5)}$$

To solve (A.5) let us put

$$\dot{\Phi} = u(\Phi)^{1/2} \qquad \text{(A.6)}$$

Therefore

$$\ddot{\Phi} = \frac{1}{2}u^{-1/2}u'(\Phi)\dot{\Phi} = \frac{1}{2}u^{-1/2}u'(\Phi)u^{1/2} \qquad \text{using (A.6)}$$

So that

$$\ddot{\Phi} = \frac{1}{2}u'(\Phi) \qquad \text{(A.7)}$$

Then from (A.5), using (A.6) and (A.7) one finds

$$\frac{1}{4}u'^2 + V'^2 + u'V' - \frac{3}{2}u^2 - 3Vu = 0$$

Which simplifies to

$$u'^2 + 4V'^2 + 4u'V' - 6u^2 - 12uV = 0 \qquad \text{(A.8)}$$

Here $u' = du/d\Phi$ and $V' = dv/d\Phi$

A solution of (A.8) is

$$\mathbf{u} = -2\mathbf{V} \qquad \text{(A.9)}$$

One can check this by observing from (A.9) that

$$\mathbf{u}' = -2\mathbf{V}' \qquad \text{(A.10)}$$

When (A.9) and (A.10) is substituted in (A.8) the result is verified.

Therefore the conclusion is that $\mathbf{u} = -2\mathbf{V}$ is a solution of (A.8)

However the solution $\mathbf{u} = -2\mathbf{V}$ is rejected because when this solution is substituted in (A.3), we obtain a static universe i.e. $\dot{\mathbf{a}} = \mathbf{0}$ so that $\mathbf{a}$ = constant.

So to obtain a sensible solution of (A.8) let us assume

$$\mathbf{u} = -2\mathbf{V} + \boldsymbol{\theta}(\boldsymbol{\Phi}) \qquad \text{(A.11)}$$

Where $\boldsymbol{\theta}(\boldsymbol{\Phi})$ is an arbitrary function of $\Phi$.

Substituting (A.11) into (A.8) one gets

$$(-2\mathbf{V}' + \boldsymbol{\theta}')^2 + 4\mathbf{V}'^2 + 4(-2\mathbf{V}' + \boldsymbol{\theta}')\mathbf{V}' - 6(-2\mathbf{V} + \boldsymbol{\theta})^2 - 12(-2\mathbf{V} + \boldsymbol{\theta})\mathbf{V} = \mathbf{0}$$

Which after simplification yields

$$\boldsymbol{\theta}'^2 + 12\mathbf{V}\boldsymbol{\theta} - 6\boldsymbol{\theta}^2 = \mathbf{0} \qquad \text{(A.12)}$$

From (A.12) we find

$$\mathbf{V} = \mathbf{V}(\boldsymbol{\Phi}) = \frac{6\boldsymbol{\theta}^2 - \boldsymbol{\theta}'^2}{12\boldsymbol{\theta}} = \left(\frac{\boldsymbol{\theta}}{2} - \frac{\boldsymbol{\theta}'^2}{12\boldsymbol{\theta}}\right) \qquad \text{(A.13)}$$

Hence we conclude that (A.11) is the solution of (A.8) i.e. solutions of (A.1) and (A.2) if $\mathbf{V}(\boldsymbol{\Phi})$ is given by (A.13).It is to be noted that (A.8) is the consequence of (A.1) and (A.2)

The function $\boldsymbol{\theta}(\boldsymbol{\Phi})$ is of course arbitrary.

Now we find from (A.3)

$$\frac{\dot{\mathbf{a}}}{\mathbf{a}} = \frac{1}{\sqrt{3}}\left(\frac{1}{2}\dot{\boldsymbol{\Phi}}^2 + \mathbf{V}(\boldsymbol{\Phi})\right)^{1/2}$$

$$= \frac{1}{\sqrt{6}}\left(\dot{\boldsymbol{\Phi}}^2 + 2\mathbf{V}(\boldsymbol{\Phi})\right)^{1/2}$$

$= \frac{1}{\sqrt{6}}(\mathbf{u} + 2\mathbf{V})^{1/2}$ Using (A.6)

i.e. $\frac{\dot{a}}{a} = \frac{1}{\sqrt{6}}\theta(\Phi)^{1/2}$ (A.14) using (A.11)

Now we like to calculate the scalar field potential $\mathbf{V}(\Phi)$ in terms of time t.

To do this we write

$\theta(\Phi) = \mathbf{f}(\mathbf{t})$ (A.15) since $\Phi$ depends on $\mathbf{t}$ only

And $\mathbf{V}(\Phi) = \mathbf{w}(\mathbf{t})$ (A.16)

Then (A.14) can be rewritten as

$$\frac{\dot{a}}{a} = \frac{1}{\sqrt{6}}\mathbf{f}(\mathbf{t})^{\frac{1}{2}} \quad \text{(A.16a) using (A.15)}$$

Now from (A.15) we have

$$\theta'(\Phi) = \frac{d\theta}{d\Phi} = \frac{d\theta}{dt}\frac{dt}{d\Phi} = \frac{df}{dt} \times \frac{1}{\dot{\Phi}} \quad \text{using (A.15)}$$

i.e. $\theta'(\Phi) = \dot{f}(t)/\dot{\Phi}$ (A.17)

So that from (A.6) and (A.11) one obtains

$\mathbf{u} = \dot{\Phi}^2 = -2\mathbf{V} + \theta(\Phi)$

$= -\theta + \frac{\theta'^2}{6\theta} + \theta$ using (A.13)

i.e. $\mathbf{u} = \frac{\theta'^2}{6\theta}$ (A.18)

i.e $\dot{\Phi}^2 = \frac{\dot{f}(t)^2}{\dot{\Phi}^2} \times \frac{1}{6f(t)}$ using (A.6), (A.17) & (A.15)

Therefore $\dot{\Phi}^4 = \frac{\dot{f}^2}{6f}$

Hence $\dot{\Phi}^2 = -\frac{1}{\sqrt{6}}\frac{\dot{f}}{f^{1/2}}$ (A.19) [Negative sign is considered for convenience]

Next we find from (A.13) and (A.16)

$\mathbf{V}(\Phi) = \mathbf{w}(\mathbf{t}) = \frac{\theta}{2} - \frac{\theta'^2}{12\theta}$

$= \frac{f(t)}{2} - \frac{\dot{f}^2}{\dot{\Phi}^2} \times \frac{1}{12f}$ Using (A.15) & (A.17)

$= \frac{f(t)}{2} - \frac{\dot{f}^2}{12f} \times \frac{1}{\dot{\Phi}^2}$

$= \frac{f(t)}{2} - \frac{\dot{f}^2}{12f} \times \frac{-\sqrt{6}f^{1/2}}{\dot{f}}$ Using (A.19)

i.e. $\mathbf{w}(\mathbf{t}) = \frac{f(t)}{2} + \frac{\dot{f}(t)}{2\sqrt{6}f^{1/2}}$ (A.20)

The above calculations assure that the exact solution of (A.1) and (A.2) can be found from the following prescription:

Choose an arbitrary function $\mathbf{f} = \mathbf{f(t)}$. For this arbitrary function $\mathbf{f(t)}$the exact solutions of (A.1) and (A.2) are:

$$\left.\begin{aligned} \frac{\dot{a}}{a} &= \frac{1}{\sqrt{6}}\mathbf{f(t)}^{\frac{1}{2}} \\ \mathbf{V(\Phi)} = \mathbf{w(t)} &= \frac{\mathbf{f(t)}}{2} + \frac{\dot{\mathbf{f}}(\mathbf{t})}{2\sqrt{6}\mathbf{f}^{1/2}} \\ \text{and}\ \ \dot{\Phi}^2 &= -\frac{1}{\sqrt{6}}\frac{\dot{\mathbf{f}}}{\mathbf{f}^{1/2}} \end{aligned}\right\} \quad \text{(A.21)}$$

We can choose the arbitrary function in an infinite number of ways. Hence we can get an infinite number of exact models from (A.21)

[ One can check that (A.21) is the exact solution set of (A.1) and (A.2) in the following way:

From the last of (A.21) one gets

$$2\dot{\Phi}\ddot{\Phi} = -\frac{1}{\sqrt{6}}\left[\frac{\ddot{\mathbf{f}}}{\mathbf{f}^{1/2}} - \frac{\mathbf{f}^{-3/2}}{2}\dot{\mathbf{f}}^2\right]$$

i.e. $\dot{\Phi}\ddot{\Phi} = -\frac{1}{2\sqrt{6}}\left[\frac{\ddot{\mathbf{f}}}{\mathbf{f}^{1/2}} - \frac{\mathbf{f}^{-3/2}}{2}\dot{\mathbf{f}}^2\right]$ **(A.22)**

Therefore, $\ddot{\Phi} + \frac{3\dot{a}}{a}\dot{\Phi} + \mathbf{V}'(\Phi) = \ddot{\Phi} + \frac{3\dot{a}}{a}\dot{\Phi} + \frac{d\mathbf{V}(\Phi)}{d\Phi}$

$= \ddot{\Phi} + \frac{3\dot{a}}{a}\dot{\Phi} + \frac{d\mathbf{w(t)}}{d\Phi}$ using (A.16)

$= \ddot{\Phi} + \frac{3\dot{a}}{a}\dot{\Phi} + \frac{d\mathbf{w(t)}}{dt} \times \frac{dt}{d\Phi}$

$= \ddot{\Phi} + \frac{3\dot{a}}{a}\dot{\Phi} + \frac{\dot{\mathbf{w}}(\mathbf{t})}{\dot{\Phi}}$

So that $\ddot{\Phi} + \frac{3\dot{a}}{a}\dot{\Phi} + \mathbf{V}'(\Phi) = \frac{1}{\dot{\Phi}}\left[\dot{\Phi}\ddot{\Phi} + 3\frac{\dot{a}}{a}\dot{\Phi}^2 + \dot{\mathbf{w}}(\mathbf{t})\right]$ **(A.23)**

From the 2nd of (A.21) one obtains

$$\dot{\mathbf{w}}(\mathbf{t}) = \frac{\dot{\mathbf{f}}(\mathbf{t})}{2} + \frac{\ddot{\mathbf{f}}(\mathbf{t})}{2\sqrt{6}\mathbf{f}^{1/2}} - \frac{1}{4\sqrt{6}}\mathbf{f}^{-3/2}\dot{\mathbf{f}}^2 \quad \textbf{(A.24)}$$

It is now easy to verify from (A.23) that

$$\ddot{\Phi} + 3\frac{\dot{a}}{a}\dot{\Phi} + \mathbf{V}'(\Phi) = \frac{1}{\dot{\Phi}}\left[\dot{\Phi}\ddot{\Phi} + 3\frac{\dot{a}}{a}\dot{\Phi}^2 + \dot{\mathbf{w}}(\mathbf{t})\right]$$

$$= \frac{1}{\dot{\Phi}}\left[-\frac{1}{2\sqrt{6}}\frac{\ddot{\mathbf{f}}}{\mathbf{f}^{1/2}} + \frac{1}{4\sqrt{6}}\mathbf{f}^{-3/2}\dot{\mathbf{f}}^2 + \frac{3\mathbf{f}^{1/2}}{\sqrt{6}} \times -\frac{1}{\sqrt{6}}\frac{\dot{\mathbf{f}}}{\mathbf{f}^{1/2}} + \frac{\dot{\mathbf{f}}}{2} + \frac{1}{2\sqrt{6}}\frac{\ddot{\mathbf{f}}}{\mathbf{f}^{1/2}} - \frac{1}{4\sqrt{6}}\mathbf{f}^{-3/2}\dot{\mathbf{f}}^2\right]$$

$$= \frac{1}{\dot{\Phi}}\left[-\frac{1}{2\sqrt{6}}\frac{\ddot{f}}{f^{1/2}} + \frac{1}{4\sqrt{6}}f^{-3/2}\dot{f}^2 - \frac{\dot{f}}{2} + \frac{\dot{f}}{2} + \frac{1}{2\sqrt{6}}\frac{\ddot{f}}{f^{1/2}} - \frac{1}{4\sqrt{6}}f^{-3/2}\dot{f}^2\right]$$

$= \frac{1}{\dot{\Phi}} \times 0 = 0$ using (A.22), (A.24) & A (21) and as $\dot{\Phi} \neq 0$ since $\Phi$ evolves continuously.

Finally one can check in a straight forward way from (A.1) that

$$\frac{\dot{a}^2}{a^2} = \frac{1}{3}\left(\frac{1}{2}\dot{\Phi}^2 + V(\Phi)\right) = \frac{1}{3}\left(\frac{1}{2}\dot{\Phi}^2 + w(t)\right) \qquad \text{Using (A.16)}$$

$$= \frac{1}{3}\left\{\frac{-1}{2\sqrt{6}}\frac{\dot{f}}{f^{1/2}} + \frac{f(t)}{2} + \frac{\dot{f}(t)}{2\sqrt{6}f^{1/2}}\right\} \qquad \text{Using last of (A.21) and (A.20)}$$

$$= \frac{1}{6}f(t)$$

i.e. $\dot{a}/_a = (1/_{\sqrt{6}})f(t)^{1/2}$ ]

Now we will construct an exact inflationary model from the exact solutions obtained before.

Let us choose the arbitrary function $f(t)$ as

$$f(t) = \left(\frac{A}{t} + B\right)^2 = \frac{A^2}{t^2} + \frac{2AB}{t} + B^2 \qquad \textbf{(A.25)}$$

i.e. $f^{1/2} = \frac{A}{t} + B$ **(A.26)** [taking positive sign of square root only]

It has to be remembered that $f(t)$ is arbitrary.

Then from (A.14) and (A.15).

$$\frac{\dot{a}}{a} = \frac{1}{\sqrt{6}}\theta(\Phi)^{1/2}$$

$$= \frac{1}{\sqrt{6}}f(t)^{1/2}$$

$$= \frac{1}{\sqrt{6}}\left(\frac{A}{t} + B\right) \qquad \text{using (A.26)}$$

Hence $\ln a = \frac{A\ln t}{\sqrt{6}} + \frac{Bt}{\sqrt{6}} + \ln A_o$ $\quad \ln A_o$= Constant of integration

i.e. $a = A_o t^{A/\sqrt{6}} e^{\frac{B}{\sqrt{6}}t}$ **(A.27)**

Now one finds from (A.26)

$$\frac{1}{2}f^{-1/2}\dot{f} = -\frac{A}{t^2} \quad \textbf{(A.28)}$$

Using (A.25) and (A.28) we find from (A.20)

$$w(t) = V(\Phi) = \frac{A^2}{2t^2} + \frac{AB}{t} + \frac{B^2}{2} - \frac{A}{\sqrt{6}t^2}$$

$$= \left(\frac{A^2}{2} - \frac{A}{\sqrt{6}}\right)\frac{1}{t^2} + \frac{AB}{t} + \frac{B^2}{2} \quad \textbf{(A.29)}$$

Equation (A.29) gives the time dependent form of the potential which gives the scale factor (A.27)

Next we will find the scalar field $\Phi$ dependence of the potential in the following way:

From the last of (A.21), we have

$$\dot{\Phi}^2 = -\frac{\dot{f}}{\sqrt{6}f^{1/2}} = -\frac{1}{\sqrt{6}} \times \frac{-2A}{t^2} \text{ Using (A.28)}$$

*i.e.* $\dot{\Phi}^2 = +\frac{\sqrt{2}}{\sqrt{3}}\frac{A}{t^2}$ **(A.29a)**

Therefore $\dot{\Phi} = -\frac{A_1}{t}$ (A.30)

Where $A_1 = \left(\sqrt{\frac{2}{3}}A\right)^{1/2}$ and negative sign is taken for convenience.

So that from (A.30) one obtains

$\Phi = -A_1 \ln t + A_2$ (A.31) where $A_2$=constant of integration.

From (A.31) we find $\ln t = \frac{A_2 - \Phi}{A_1}$

i.e. $t = e^{(A_2 - \Phi)/A_1}$

therefore $\frac{1}{t} = e^{-(A_2-\Phi)/A_1} = K_o e^{\Phi/A_1}$ (A.32)

Where $K_o = e^{-A_2/A_1}$ = constant

Now using (A.32), we find from (A.29)

$$V(\Phi) = \left(\frac{A^2}{2} - \frac{A}{\sqrt{6}}\right) K_o^2\, e^{2\Phi/A_1} + AB.K_o e^{\Phi/A_1} + \frac{B^2}{2} \quad \text{(A.33)}$$

Equation (A.33) gives the $\mathbf{\Phi}$ dependence of the potential which in more compact form can be recasted as

$$\mathbf{V(\Phi) = C.e^{2\Phi/A_1} + D.e^{\Phi/A_1} + \frac{B^2}{2}} \qquad \text{(A.34)}$$

Where $\mathbf{C} = \left(\frac{\mathbf{A^2}}{\mathbf{2}} - \frac{\mathbf{A}}{\sqrt{\mathbf{6}}}\right)\mathbf{K_o^2}$ = constant

and $\mathbf{D} = \mathbf{ABK_o}$ = Constant

Thus it turns out that the potential given by (A.34) produces the scale factor given by (A.27). The potential given by (A.34) and (A.29) are the same potential in different forms.

References:-